\documentstyle[11pt,newpasp,twoside]{article}
\def\edcomment#1{\iffalse\marginpar{\raggedright\sl#1\/}\else\relax\fi}
\reversemarginpar

\begin{document}
\title{Multiwavelength Observations of Galactic Winds: \\ Near and Far}
 \author{Sylvain Veilleux\altaffilmark{1,2}}
\affil{Department of Astronomy, University of Maryland, College Park, MD 20742}

\altaffiltext{1}{Current Address: 320-47 Downs Lab., Caltech, Pasadena,
CA 91125 and Observatories of the Carnegie Institution of Washington,
813 Santa Barbara Street, Pasadena, CA 91101}

\altaffiltext{2}{Cottrell Scholar of the Research Corporation}

\begin{abstract}
This paper provides a critical discussion of the observational
evidence for winds in our own Galaxy, in nearby star-forming and
active galaxies, and in the high-redshift universe. The implications
of galactic winds on the formation and evolution of galaxies and the
intergalactic medium are briefly discussed. A number of observational
challenges are mentioned to inspire future research directions.
\end{abstract}

\section{Introduction}
Active galactic nuclei (AGN) and nuclear starbursts may severely
disrupt the gas phase of galaxies through deposition of a large amount
of mechanical energy in the centers of galaxies.  As a result, a
large-scale galactic wind that encompasses much of the central regions
of these galaxies may be created (e.g., Chevalier \& Clegg 1985;
Schiano 1985).  Depending upon the extent of the gaseous halo and its
density and upon the wind's mechanical luminosity and duration, the
wind may ultimately blow out through the halo and into the
intergalactic medium.  The effects of these winds may be
far-reaching. Bregman (1978) has suggested that the Hubble sequence
can be understood in terms of a galaxy's greater ability to sustain
winds with increasing bulge-to-disk ratio. Galactic winds may affect
the thermal and chemical evolution of galaxies and the intergalactic
medium by depositing large quantities of hot, metal-enriched material
on the outskirts of galaxies and beyond. This widespread circulation
of matter and energy between the disks and halos of galaxies may be
responsible for the mass-metallicity relation between galaxies.

This paper reviews the observed properties (\S 2) and impact (\S 3) of
starburst- and AGN-driven winds in both the local and distant
universe; the discussion on starburst-driven winds is largely borrowed
from Veilleux (2003), while new elements on AGN outflows are also
included. The last section (\S 4) discusses future avenues of
research. The theory and numerical modelling of galactic winds are not
discussed here due to space limitations (see, e.g., Veilleux et
al. 2002; Strickland 2002; Heckman 2002; Veilleux 2003).  Collimated
jet outflows and unresolved nuclear winds in AGNs are also beyond the
scope of this paper; recent reviews of these topics include Zensus
(1997), Veilleux et al. (2002), and Crenshaw, Kraemer, \& George
(2003).

\section{Observed Properties of Galactic Winds}

AGN- and starburst-driven winds are common among local galaxies.
Galaxies with global star formation rates per unit area $\Sigma_*
\equiv SFR / \pi R_{\rm opt}^2 \ga 0.1$ M$_\odot$ yr$^{-1}$
kpc$^{-2}$, where $R_{\rm opt}$ is the optical radius, often show
signs of large-scale winds. This general rule-of-thumb also appears to
apply to ultra/luminous infrared galaxies (see \S 2.3) and distant
Lyman break galaxies (see \S 2.4). ``Quiescent'' galaxies with lower
star formation rates per unit area often show thick ionized disks, but
no galactic-scale outflow (e.g., Miller \& Veilleux 2003a,
2003b). This rule-of-thumb is conservative since a number of known
starburst-driven wind galaxies, including our own Galaxy (\S 2.1) and
several dwarf galaxies, have $\Sigma_* <<$ 0.1 M$_\odot$ yr$^{-1}$
kpc$^{-2}$ (e.g., Hunter \& Gallagher 1990, 1997; Meurer et al. 1992;
Marlowe et al. 1995; Kunth et al. 1998; Martin 1998, 1999). The
production of detectable winds depends not only on the characteristics
of the energy source (AGN vs starburst, power, age), but also on the
detailed properties of the ISM in the host galaxies (e.g., see the
theoretical blowout criterion of MacLow \& McCray 1988).
% and MacLow, McCray, \& Norman 1989).

The winds in active and star-forming galaxies in the local universe
show a very broad range of properties, with opening angles of $\sim$
0.1 -- 0.5 $\times$ (4$\pi$ sr), radii ranging from $<$ 1 kpc to
several 10s of kpc, outflow velocities of a few 10s of km s$^{-1}$ to
more than 1000 km s$^{-1}$ (with clear evidence for a positive
correlation with the temperature of the gas phase), total (kinetic and
thermal) outflow energies of $\sim$ 10$^{53}$ -- 10$^{57}$ ergs, and
mass outflow rates ranging from $<$ 1 M$_\odot$ yr$^{-1}$ to $>$ 100
M$_\odot$ yr$^{-1}$. In AGNs, the mass outflow rates on kpc scale are
larger than the mass accretion rates needed to power the central
nucleus. In starbursts, the mass outflow rates scale roughly with the star
formation rates (see \S 2.3 below).

In the remainder of this section, we repeat the discussion of
Veilleux (2003) on a few well-studied cases of galactic winds in the
local universe and summarize the evidence for winds in luminous and
ultraluminous infrared galaxies at low and moderate redshifts as well
as in distant Lyman break galaxies.

\subsection{The Milky Way}

By far the closest case of a large-scale outflow is the wind in our
own Galaxy. Evidence for a dusty bipolar wind extending $\sim$ 350 pc
($\sim$ 1$^\circ$) above and below the disk of our Galaxy has recently
been reported by Bland-Hawthorn \& Cohen (2003) based on data from
the Midcourse Space Experiment (MSX). The position of the warm dust
structure coincides closely with the well-known Galactic Center Lobe
detected at radio wavelengths (e.g., Sofue 2000 and references
therein). Simple arguments suggest that the energy requirement for
this structure is of order $\sim$ 10$^{55}$ ergs with a dynamical time
scale of $\sim$ 1 Myr.

Bland-Hawthorn \& Cohen (2003) also argue that the North Polar Spur, a
thermal X-ray/radio loop that extends from the Galactic plane to $b =
+80^\circ$ (e.g., Sofue 2000), can naturally be explained as an
open-ended bipolar wind, when viewed in projection in the near
field. This structure extends on a scale of 10 -- 20 kpc and implies
an energy requirement of $\sim$ (1 -- 30) $\times$ 10$^{55}$ ergs and
a dynamical timescale of $\sim$ 15 Myr, i.e.  considerably longer than
that of the smaller structure seen in the MSX maps.  If confirmed,
this may indicate that the Milky Way Galaxy has gone through multiple
galactic wind episodes.  Bland-Hawthorn \& Cohen (2003) point out that
the North Polar Spur would escape detection in external galaxies; it
is therefore possible that the number of galaxies with large-scale
winds has been (severely?) underestimated.

\subsection{Nearby Starburst Galaxies}

Two classic examples of starburst-driven outflows are described in
this section to illustrate the wide variety of processes taking place
in these objects.

\vskip 0.1in
\noindent{\bf M~82.}  This archetype starburst galaxy hosts arguably
the best studied galactic wind. Some of the strongest evidence for the
wind is found at optical wavelengths, where long-slit and Fabry-Perot
spectroscopy of the warm ionized filaments above and below the disk
shows line splittings of up to $\sim$ 250 km s$^{-1}$, corresponding
to deprojected velocities of order 525 -- 655 km s$^{-1}$ (e.g.,
%McCarthy, Heckman, \& van Breugel 1987; Bland \& Tully 1988; 
McKeith et al. 1995; Shopbell \& Bland-Hawthorn 1998). Combining these
velocities with estimates for the ionized masses of the outflowing
filamentary complex, the kinetic energy involved in the warm ionized
outflow is of order $\sim$ 2 $\times$ 10$^{55}$ ergs or $\sim$ 1\% of
the total mechanical energy input from the starburst.  The ionized
filaments are found to lie on the surface of cones with relatively
narrow opening angles ($\sim$ 5 -- 25$^\circ$) slightly tilted ($\sim$
5 -- 15$^\circ$) with respect to the spin axis of the galaxy. Deep
narrow-band images of M82 have shown that the outflow extends out to
at least 12 kpc on one side (e.g., Devine \& Bally 1999), coincident
with X-ray emitting material seen by $ROSAT$ (Lehnert, Heckman, \&
Weaver 1999) and $XMM$-Newton (Stevens, Read, \& Bravo-Guerrero
2003). The wind fluid in this object has apparently been detected by
both $CXO$ (Griffiths et al. 2000) and $XMM$-Newton (Stevens et
al. 2003). The well-known H~I complex around this system (e.g., Yun et
al. 1994) may be taking part, and perhaps even focussing, the outflow
on scales of a few kpc (Stevens et al. 2003). Recently published
high-quality CO maps of this object now indicate that some of the
molecular material in this system is also involved in the large-scale
outflow (Walter, Weiss, \& Scoville 2002; see also Garcia-Burillo et
al. 2001). The outflow velocities derived from the CO data ($\sim$ 100
km s$^{-1}$ on average) are considerably lower than the velocities of
the warm ionized gas, but the mass involved in the molecular outflow
is substantially larger ($\sim$ 3 x 10$^8$ M$_\odot$), implying a
kinetic energy ($\sim$ 3 $\times$ 10$^{55}$ ergs) that is comparable
if not larger than that involved in the warm ionized filaments. The
molecular gas is clearly a very important dynamical component of this
outflow.

\vskip 0.1in
\noindent{\bf NGC~3079. } An outstanding example of starburst-driven
superbubble is present in the edge-on disk galaxy, NGC~3079.
High-resolution $HST$ H$\alpha$ maps of this object show that the
bubble is made of four separate bundles of ionized filaments (Cecil et
al. 2001). The two-dimensional velocity field of the ionized bubble
material derived from Fabry-Perot data (Veilleux et al. 1994)
indicates that the ionized bubble material is entrained in a mushroom
vortex above the disk with velocities of up to $\sim$ 1500 km s$^{-1}$
(Cecil et al. 2001). A recently published X-ray map obtained with the
$CXO$ (Cecil, Bland-Hawthorn, \& Veilleux 2002) reveals excellent
spatial correlation between the hot X-ray emitting gas and the warm
optical line-emitting material of the bubble, suggesting that the
X-rays are being emitted either as upstream, standoff bow shocks or by
cooling at cloud/wind conductive interfaces. This good spatial
correlation between the hot and warm gas phases appears to be common
in galactic winds (Strickland et al. 2000, 2002; Veilleux et al. 2003,
and references therein). The total energy involved in the outflow of
NGC~3079 appears to be slightly smaller than that in M~82, although it
is a lower limit since the total extent of the X-ray emitting material
beyond the nuclear bubble of NGC~3079 is not well constrained (Cecil
et al. 2002). Contrary to M~82, the hot wind fluid that drives the
outflow in NGC~3079 has not yet been detected, and evidence for
entrained molecular gas is sparse and controversial (e.g., Irwin \&
Sofue 1992; Baan \& Irwin 1995; Israel et al. 1998; but see Koda et
al. 2002).

\subsection{Luminous and Ultraluminous Infrared Galaxies.}

Given that the far-infrared energy output of a (dusty) galaxy is a
direct measure of its star formation rate, it is not surprising {\em a
posteriori} to find evidence for large-scale galactic winds in several
starburst-dominated luminous and ultraluminous infrared galaxies
(LIRGs and ULIRGs; e.g., Heckman et al. 1990; Veilleux et al. 1995).
Systematic searches for winds have been carried out in recent years in
these objects to look for the unambiguous wind signature of
blueshifted absorbing material in front of the continuum source
(Heckman et al. 2000; Rupke et al.  2002). The feature of choice to
search for outflowing neutral material in galaxies of moderate
redshifts ($z \la$ 0.6) is the Na~ID interstellar absorption doublet
at 5890, 5896 \AA. The wind detection frequency derived from a set of
44 starburst-dominated LIRGs and ULIRGs is high, of order $\sim$ 70 --
80\% (Rupke et al. 2002, 2003 in prep.). The outflow velocities reach
values in excess of 1700 km s$^{-1}$ (even more extreme velocities are
found in some AGN-dominated ULIRGs; e.g., Mrk~231; Rupke et al. 2002).

A simple model of a mass-conserving free wind (details of the model
are given in Rupke et al. 2002) is used to infer mass outflow rates in
the range $\dot{M}_{\mathrm{tot}}$(H)$\;= {\mathrm few} - 120\;$ for
galaxies hosting a wind.  These values of $\dot{M}_{\mathrm{tot}}$,
normalized to the corresponding global star formation rates inferred
from infrared luminosities, are in the range $\eta \equiv
\dot{M}_{\mathrm{tot}} / \mathrm{SFR} = 0.01 - 1$. The parameter
$\eta$, often called the ``mass entrainment efficiency'' or
``reheating efficiency'' shows no dependence on the mass of the host
(parameterized by host galaxy kinematics and absolute $R$- and
$K^{\prime}$-band magnitudes), but there is a possible tendency for
$\eta$ to decrease with increasing infrared luminosities (i.e. star
formation rates). The large molecular gas content in ULIRGs may impede
the formation of large-scale winds and reduce $\eta$ in these objects.
A lower thermalization efficiency (i.e. higher radiative efficiency)
in these dense gas-rich systems may also help explain the lower $\eta$
(Rupke et al. 2003, in prep.). 

\subsection{Lyman Break Galaxies}

Evidence for galactic winds has now been found in a number of z $\sim$
3 -- 5 galaxies, including an important fraction of Lyman break
galaxies (LBGs; e.g., Franx et al. 1997; Pettini et al. 2000, 2002;
Frye, Broadhurst, \& Benitez 2002; Dawson et al. 2002; Ajiki et
al. 2002; Adelberger et al. 2003; Shapley et al. 2003). The best
studied wind at high redshift is that of the gravitationally lensed
LBG MS~1512-cB58 (Pettini et al. 2000, 2002). An outflow velocity of
$\sim$ 255 km s$^{-1}$ is derived in this object, based on the
positions of the low-ionization absorption lines relative to the
rest-frame optical emission lines (Ly$\alpha$ is to be avoided for
this purpose since resonant scattering and selective dust absorption
of the Ly$\alpha$ photons may severely distort the profile of this
line; e.g., Tenorio-Tagle et al. 1999). The mass-conserving free wind
model of Rupke et al. (2002) applied to MS~1512-cB58 (for consistency)
results in a mass outflow rate of $\sim$ 20 $M_\odot$ yr$^{-1}$,
equivalent to about 50\% the star formation rate of this galaxy based
on the dust-corrected UV continuum level. Similar outflow velocities
are derived in other LBGs (Pettini et al. 2001). The possibly strong
impact of these LBG winds on the environment at high $z$ is discussed
in the next section (\S 3.2).

\section{Impact of Galactic Winds on the Environment}

Due to space limitations, it is not possible to discuss in detail the
profound influence of galactic winds on galaxy formation and evolution
and on the properties of the intergalactic medium. This section
reviews a few key results on the heating and enrichment of the ISM and
IGM, and describe new optical constraints on the size of the zone of
influence of galactic winds.

\subsection{Heating and Enrichment of the ISM and IGM}

\noindent{\bf Hot Metal-Enriched Gas in Starburst-Driven Winds.}
Nuclear starbursts inject both mechanical energy and metals in the
centers of galaxies. This hot, chemically-enriched material is
eventually deposited on the outskirts of the host galaxies, and
contributes to the heating and metal enrichment of galaxy halos and
the IGM. Surprisingly little evidence exists for the presence of this
enriched wind fluid. This is due to the fact that the wind fluid is
tenuous and hot and therefore very hard to detect in the X-rays. The
current best evidence for the existence of the wind fluid is found in
M~82 (Griffiths et al. 2000; Stevens et al. 2003), NGC~1569 (Martin,
Kobulnicky, \& Heckman 2002), and possibly the Milky Way (e.g., Koyama
et al. 1989; Yamauchi et al. 1990). The ratio of alpha elements to
iron appears to be slightly super-solar in the winds of both NGC~1569
and M~82, as expected if the stellar ejecta from SNe II are providing
some, but not all of the wind fluid.

\vskip 0.1in
\noindent{\bf Selective Loss of Metals in Starburst-Driven Winds. }
The outflow velocities in starburst-dominated LIRGs and ULIRGs do not
appear to be correlated with the rotation velocity (or equivalently,
the escape velocity) of the host galaxy, implying selective loss of
metal-enriched gas from shallower potentials (Heckman et al. 2000;
Rupke et al. 2002).  If confirmed over a broader range of galaxy
masses (e.g., Martin 1999; but see Martin 2003 and Rupke et al. 2003,
in prep.), this result may help explain the mass-metallicity relation
and radial metallicity gradients in elliptical galaxies and galaxy
bulges and disks (e.g., Bender, Burstein, \& Faber 1993; Franx \&
Illingworth 1990; Carollo \& Danziger 1994; Zaritsky et al. 1994;
Trager et al. 1998).
%; Jablonka et al. 1996; Jorgensen et al. 1996; Pahre et al. 1998;
%Several models have assumed that this was the case (e.g., Wyse \& Silk
%1985; Dekel \& Silk 1986; Lynden-Bell 1992; Kauffmann \& Charlot 1998
%and many others since).
The ejected gas may also contribute to the heating and chemical
enrichment of the ICM in galaxy clusters (e.g., Dupke \& Arnaud 2001;
Finoguenov et al. 2002, and references therein).
%Finoguenov, Arnaud, \& David 2001; Tamura et al. 2001

\vskip 0.1in
\noindent{\bf Heating by AGN-Driven Outflows. } The large ``cavities''
in the X-ray surface brightness of several cooling flow clusters with
radio-loud cD galaxies (e.g., 
%Fabian et al. 1981; Branduardi-Raymont et al. 1981; 
B\"ohringer et al. 1993; Fabian et al. 2000; McNamara et al. 2000,
2001;
%Fabian 2001;  Allen et al. 2001; 
David et al. 2001;
%Peterson et al. 2001; 
Heinz et al. 2002) point to the direct influence of AGN-driven
outflows on the ICM. The hot/relativistic buoyant gas injected into
the ICM by the AGN reduces and perhaps even quenches the mass accretion rates
associated with the cooling flows, possibly through thermal conduction
or ``effervescent'' heating (e.g., Quilis, Bower, \& Balogh 2001;
%Jones et al. 2002 starburst-driven outflow impacting on ICM
Churazov et al. 2002; Ruszkowski \& Begelman 2003; see Kim \& Narayan
2003, however).
%; Begelman 2003 Carnegie review...
%see turbulent mixing: Kim & Narayan 2003, astro-ph/0308376

\vskip 0.1in
\noindent{\bf Dust Outflows.}  Galactic winds also act as conveyor
belts for the dust in the hosts. The evidence for a large-scale dusty
outflow in our own Galaxy has already been mentioned in \S 2.1
(Bland-Hawthorn \& Cohen 2003). Far-infrared maps of external galaxies
with known galactic winds show extended dust emission along the galaxy
minor axis, suggestive of dust entrainment in the outflow (e.g., Hughes,
Gear, \& Robson 1994; Alton et al. 1998, 1999;
%; Alton, Davies, \& Bianchi 1999; 
Radovich, Kahanp\"a\"a, \& Lemke 2001).
% NGC 253 H_2: Sugai et al. 2003). 
Direct evidence is also found at optical wavelengths in the form of
elevated dust filaments in a few galaxies (e.g., NGC~1808, Phillips
1993; NGC~3079, Cecil et al. 2001).
%NGC~891, Howk \& Savage 1997).
A strong correlation between color excesses, $E(B - V)$, and the
equivalent widths of the blueshifted low-ionization lines in
star-forming galaxies at low (e.g., Armus, Heckman, \& Miley 1989;
Veilleux et al. 1995; Heckman et al. 2000; Rupke et al. 2003) and
moderate-to-high redshifts (e.g., Rupke et al. 2003; Shapley et
al. 2003) provides additional support for the prevalence of dust
outflows. Assuming a Galactic dust-to-gas ratio, Heckman et al.
(2000) estimate that the dust outflow rate is about 1\% of the total
mass outflow rate in LIRGs. Dust ejected from galaxies may help feed
the reservoir of intergalactic dust (e.g., Coma cluster; Stickel et
al. 1998; note, however, that tidal and ram-pressure stripping may be
more efficient than winds at carrying dust into the ICM; see
contribution by Stickel at this conference).

\subsection{Zone of Influence of Winds}

The impact of galactic winds on the host galaxies and the environment
depends sensitively on the size of the ``zone of influence'' of these
winds, i.e.  the region affected either directly (e.g., heating,
metals) or indirectly (e.g., ionizing radiation) by these winds. But
the true extent of galactic winds is often difficult to determine in
practice due to the steeply declining density profile of both the wind
material and the host ISM. The zone of influence of galactic winds is
therefore often estimated using indirect means which rely on a number
of assumptions.  

A popular method is to use the measured velocity of the outflow and
compare it with the local escape velocity derived from some model for
the gravitational potential of the host galaxy. If the measured
outflow velocity exceeds the predicted escape velocity {\em and} if
the halo drag is negligible, then the outflowing material is presumed
to escape the host galaxy and be deposited in the IGM on scales $\ga$
50 -- 100 kpc (see, e.g., Rupke et al. 2002 for an application of this
method). Another method is to rely on the expected terminal velocity
of an adiabatic wind at the measured X-ray temperature $T_X$ [$v_X
\sim (5KT_X/\mu)^{0.5}$, where $\mu$ is the mean mass per particle] to
provide a lower limit to the velocity of the wind fluid (this is a
lower limit because it only takes into account the thermal energy of
this gas and neglects any bulk motion; e.g., Chevalier \& Clegg 1985;
Martin 1999; Heckman et al. 2000).  Both of these methods make the
important assumption that halo drag is negligible. Silich \&
Tenorio-Tagle (2001) have argued that halo drag may severely limit the
extent of the wind and the escape fraction.  Drag by a dense halo or a
complex of tidal debris may be particularly important in ULIRGs if
they are created by galaxy interactions (e.g., Veilleux, Kim, \&
Sanders 2002b).

The large uncertainties on these indirect estimates of the zone of
influence of galactic winds emphasize the need for more direct
measurements; these are discussed next.

\vskip 0.1in
\noindent{\bf Deep Multiwavelength Maps of Local Galaxies.}  The
fundamental limitation in directly measuring the zone of influence of
winds is the sensitivity of the instruments. Fortunately, $CXO$ and
$XMM$-Newton now provide powerful tools to better constrain the extent
of the hot medium (e.g., M~82, Stevens et al. 2003; NGC~3079, Cecil et
al. 2002; NGC~6240, Komossa et al. 2003; Veilleux et al. 2003;
NGC~1511, Dahlem et 2003). The reader should refer to the contribution
of M. Ehle at this conference for a summary of recent X-ray results
(see also Strickland et al. 2003 and references
therein). Technological advancements have also allowed to detect
galactic winds on very large scales at radio wavelengths (e.g., Irwin
\& Saikia 2003). A discussion of these results is beyond the scope of
this short review.

The present discussion focusses on optical constraints derived from
the detection of warm ionized gas on the outskirts of wind
hosts. Progress in this area of research has been possible thanks to
advances in the fabrication of low-order Fabry-Perot etalons which are
used as tunable filters to provide monochromatic images over a large
fraction of the field of view of the imager. The central wavelength
(3500 \AA\ -- 1.0 $\mu$m) is tuned to the emission-line feature of
interest and the bandwidth (10 -- 100 \AA) is chosen to minimize the
sky background. The data acquisition methods used to reach very faint
flux limits are discussed in Veilleux (2003) and references therein,
and are not repeated here. The Taurus Tunable Filter (TTF;
Bland-Hawthorn \& Jones 1998; Bland-Hawthorn \& Kedziora-Chudczer
2003) has been used on the AAT and WHT to produce emission-line images
of several ``quiescent'' disk galaxies (Miller \& Veilleux 2003a) and
a few starburst galaxies (Veilleux et al. 2003) down to unprecedented
emission-line flux levels. 

Gaseous complexes or filaments larger than $\sim$ 20 kpc have been
discovered or confirmed in a number of wind hosts (e.g., NGC~1482 and
NGC~6240; the presence of warm ionized gas at $\sim$ 12 kpc from the
center of M~82 was discussed in \S 2.2).  Multi-line imaging and
long-slit spectroscopy of the gas found on large scale reveal line
ratios which are generally not H~II region-like.  Shocks often
contribute significantly to the ionization of the outflowing gas on
the outskirts of starburst galaxies. As expected from shock models
(e.g., Dopita \& Sutherland 1995), the importance of shocks over
photoionization by OB stars appears to scale with the velocity of the
outflowing gas (e.g., NGC~1482, or ESO484-G036 versus NGC~1705;
NGC~3079 is an extreme example of a shock-excited wind nebula;
Veilleux et al. 1994), although other factors like the starburst age,
star formation rate, and the dynamical state of the outflowing
structure (e.g., pre- or post-blowout) must also be important in
determining the excitation properties of the gas at these large radii
(e.g., Shopbell \& Bland-Hawthorn 1998 and Veilleux~\&~Rupke~2002). In
the cases of AGN-driven winds, the hard radiation from the central
source sometimes produces highly ionized winds and/or large-scale
ionization cones (e.g., NGC~1068, NGC~1365 and NGC~4388; Veilleux et
al. 2003).

%\vskip 0.1in
%\noindent{\bf Influence of the Wind on Companion Galaxies.}  Companion
%galaxies located within the zone of influence of the wind will be
%affected by the wind ram pressure. Irwin et al. (1987) noticed that
%the dwarf S0 galaxy NGC~3073 exhibits an elongated H~I tail that is
%remarkably aligned with the nucleus of NGC~3079. Irwin et al. have
%argued that ram pressure due to the outflowing gas of NGC~3079 is
%responsible for this tail. If that is the case, the wind of NGC~3079
%must extent to at least $\sim$ 50 kpc. This is the only system known
%so far where this phenomenon is suspected to take place.

\vskip 0.1in
\noindent{\bf Lyman Break Galaxies.}  Large absorption-line data sets
collected on high-$z$ galaxies provide new constraints on the zone of
influence of winds in the early universe. Adelberger et al. (2003)
have recently presented tantalizing evidence for a deficit of neutral
hydrogen clouds within a comoving radius of $\sim$ 0.5 $h^{-1}$ Mpc
from $z \sim 3$ LBGs. The uncertainties are large and the results are
significant at less than the $\sim$ 2$\sigma$ level. Adelberger et
al. (2003) argue that this deficit, if real, is unlikely to be due
solely to the ionizing radiation from LBGs (e.g., Steidel et al. 2001;
Giallongo et al. 2002). They favor a scenario in which the winds in
LBGs directly influence the surrounding IGM. They also argue that the
excess of absorption-line systems with large CIV column densities near
LBGs is evidence for chemical enrichment of the IGM by the LBG winds.

\section{Future Avenues of Research}

Although great strides have been made over the last decade in
understanding the physics and impact of galactic winds in the local
and distant universe, much work remains to be done to be able to
quantify the overall role of these winds on the formation and
evolution of galaxy-sized structures. Absorption-line studies of
bright background galaxies (e.g., high-$z$ quasars, LBGs) have proven
to be a very powerful tool to constrain the zone of influence of
galactic winds at large redshifts. The next generation of instruments
on $HST$ will provide the capabilities to extend the sample to a
larger set of wind galaxies.  $CXO$ and $XMM$-Newton will continue
their harvest of high-quality data on the hot medium in galactic
winds, and within five years a new generation of radio telescopes
(e.g., $EVLA$, $SKA$, $CARMA$, $ALMA$) will probe the hot relativistic
component of galactic winds better than ever before and provide the
sensitivity to better quantify the role of the molecular gas in the
dynamics of local winds. This component may be particularly important
in determining the overall thermalization efficiency of galactic
winds, or the percentage of the mechanical energy from the starburst
or AGN that goes into heating the gas and driving the outflow; this
quantity is currently very poorly constrained.  The advent of tunable
filters on 8-meter class telescopes [e.g., OSIRIS on the GranTeCan
(Cepa et al. 2000) and the Maryland-Magellan Tunable Filter on the
Baade 6.5-m telescope] should improve the sensitivity of optical wind
surveys at least tenfold.  Measurements with this second generation of
tunable filters will provide direct quantitative constraints on the
gaseous cross-section of active and star-forming galaxies, and the
importance of mass exchange between galaxies and their
environment. These powerful instruments will be ideally suited to
search for galaxies with starburst-driven winds, exploiting the
contrast in the excitation properties of the wind component and the
star-forming disk (Veilleux \& Rupke 2002).

\acknowledgements

Special thanks to P.-A. Duc for organizing an excellent
conference. Some of the results presented in this paper are part of a
long-term effort involving many collaborators, including
J. Bland-Hawthorn, G. Cecil, P. L. Shopbell, and R. B. Tully and
Maryland graduate students S. T. Miller and D. S. Rupke. This article
was written while the author was on sabbatical at the California
Institute of Technology and the Observatories of the Carnegie
Institution of Washington; the author thanks both of these
institutions for their hospitality.  The author acknowledges partial
support of this research by a Cottrell Scholarship awarded by the
Research Corporation, NASA/LTSA grant NAG 56547, and NSF/CAREER grant
AST-9874973.

%\reference Dupond, H. 1999, \apj, 41, 432  
%\reference Durant, M., \& Martin, D. 2004, \aap, 4324, 5453 

\end{document}